\documentclass[9pt,twocolumn,twoside]{osajnl}

\journal{josaa} 

\setboolean{shortarticle}{false} 

\usepackage{supertabular}
\usepackage{dblfloatfix} 
    
\begin{document}

\title{Comparative analysis of imaging configurations and objectives for Fourier microscopy}

\author[1]{Jonathan A. Kurvits}
\author[1]{Mingming Jiang}
\author[1,*]{Rashid Zia}

\affil[1]{School of Engineering and Department of Physics, Brown University, Providence, RI 02912}

\affil[*]{Corresponding author: rashid\_zia@brown.edu}

\dates{Compiled \today}

\ociscodes{(070.0070) Fourier optics and signal processing; (220.3620) Lens system design; (110.0180) Microscopy; (180.2520) Fluorescence microscopy.}

\doi{\copyright~2015 Optical Society of America. One print or electronic copy may be made for personal use only. Systematic reproduction and distribution, duplication of any material in this paper for a fee or for commercial purposes, or modifications of the content of this paper are prohibited. (Link to published abstract and article: \url{http://dx.doi.org/10.1364/JOSAA.32.002082})}

\begin{abstract}
Fourier microscopy is becoming an increasingly important tool for the analysis of optical nanostructures and quantum emitters. However, achieving quantitative Fourier space measurements requires a thorough understanding of the impact of aberrations introduced by optical microscopes, which have been optimized for conventional real-space imaging. Here, we present a detailed framework for analyzing the performance of microscope objectives for several common Fourier imaging configurations. To this end, we model objectives from Nikon, Olympus, and Zeiss using parameters that were inferred from patent literature and confirmed, where possible, by physical disassembly. We then examine the aberrations most relevant to Fourier microscopy, including the alignment tolerances of apodization factors for different objective classes, the effect of magnification on the modulation transfer function, and vignetting-induced reductions of the effective numerical aperture for wide-field measurements. Based on this analysis, we identify an optimal objective class and imaging configuration for Fourier microscopy. In addition, as a resource for future studies, the Zemax files for the objectives and setups used in this analysis have been made publicly available. 
\end{abstract}

\setboolean{displaycopyright}{true}

\maketitle
\thispagestyle{fancy}
\ifthenelse{\boolean{shortarticle}}{\abscontent}{}
\renewcommand\arraystretch{1.1}
\setlength{\tabcolsep}{3pt}
\section{Introduction}
Since the work of Lieb {\it{et al.}}~\cite{lieb2004single}, Fourier microscopy has become an increasingly important experimental technique for nano-optics. It is now commonly used to study quantum emitters~\cite{lieb2004single,castelletto2010imaging,taminiau2012quantifying,karaveli2013time,schuller2013orientation,backer2013single,dodson2014wide}, optical nanostructures~\cite{drezet2008leakage,huang2008gain,bharadwaj2011electrical,sersic2011fourier,shegai2011bimetallic,shegai2011unidirectional,bernal2012plasmonic}, and the interactions of these two systems~\cite{buchler2005measuring,mattheyses2005fluorescence,tang2007effects,curto2010unidirectional,aouani2011bright,lee2011planar,shegai2011angular,shegai2012directional,wagner2012back,zhu2012direct,guebrou2012coherent,hartmann2012launching,hartmann2013radiation,wang2013directional,curto2013multipolar,hancu2013multipolar,chen2014tamm,schokker2014lasing,shi2014spatial,bulgarini2014nanowire,osorio2015kspace,mohtashami2015Angle}. For example, Fourier microscopy has been used to characterize the orientation of single molecules~\cite{lieb2004single,backer2013single} and luminescent excitons in layered materials~\cite{schuller2013orientation}, the radiation pattern and directivity of optical antennas~\cite{curto2010unidirectional,hancu2013multipolar}, and the multipolar origin of quantum transitions~\cite{taminiau2012quantifying,karaveli2013time,dodson2014wide}. These Fourier microscopy studies all share a common goal, namely to measure quantitative information about the angular spectrum radiated by a microscopic sample.

However, a surprisingly wide range of optical systems and setups have been used to achieve this goal, including many different objective classes with varying levels of aberration correction and numerical aperture (NA). For example, researchers have used everything from dry objectives with 0.8 NA~\cite{bulgarini2014nanowire} to 1.49 NA total internal reflection fluorescence (TIRF) objectives~\cite{shegai2011bimetallic,shegai2011unidirectional,shegai2012directional,guebrou2012coherent} and even 1.65 NA high-index objectives~\cite{lee2011planar}. Researchers have also used several different configurations to image the back focal plane (BFP). Some configurations place a Bertrand lens  before the microscope's image plane~\cite{taminiau2012quantifying,schuller2013orientation,karaveli2013time}; others place a Bertrand lens after the image plane~\cite{hartmann2013radiation,dodson2014wide}, while a third set use relay optics to reimage and magnify the BFP~\cite{shegai2011bimetallic,shegai2011angular,hancu2013multipolar,curto2013multipolar,curto2015personal}. 
\begin{figure*}[b]
\includegraphics{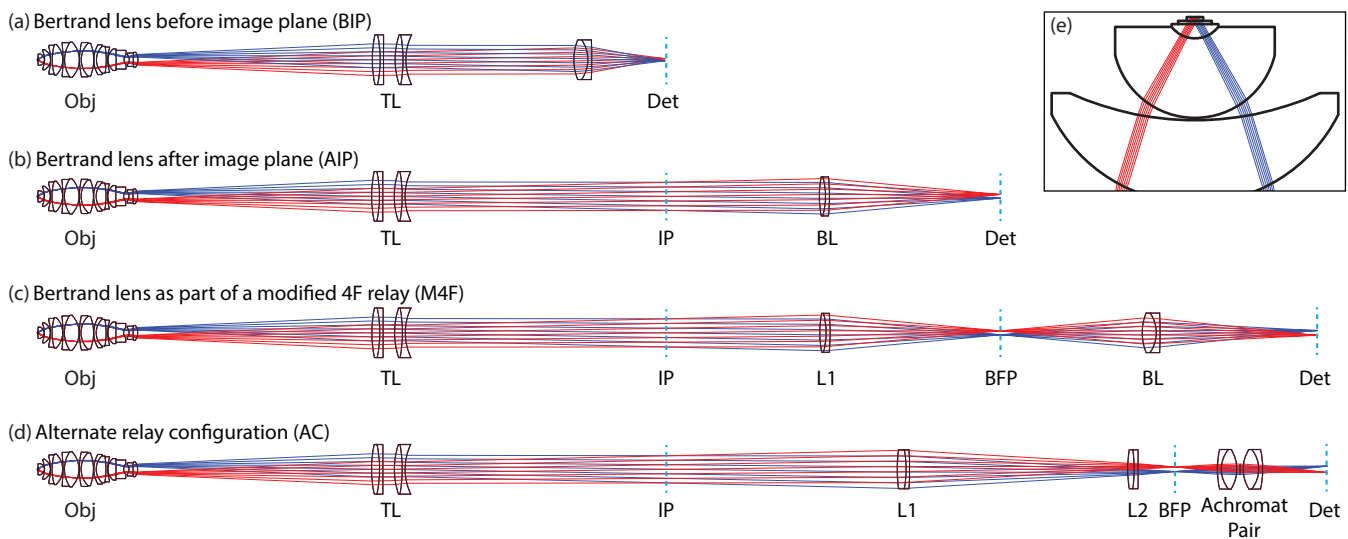}
\caption{Schematic of various Fourier imaging techniques showing relative the positions of objective (Obj), tube lens (TL), image planes (IP) and Fourier imaging optics. In (a) and (b), a single Bertrand lens (BL) is used to image the back-focal-plane (BFP) of the microscope objective onto a detector (Det). In (c), a modified 4f relay is used to magnify the BFP and re-image it further from the microscope. In (d), relay lenses L1 and L2 magnify the BFP image, which is then re-imaged onto a detector using an achromatic pair. Blue and red lines show how rays emitted at two different angles ($\pm30^\circ$) from the object focus to different locations on the detector plane. (e) Magnified view of first two elements in the objective highlights the origin of blue and red lines from two different angles.}
\label{FourierConfigs}
\end{figure*}

Beyond nano-optics, Fourier microscopy is also becoming an important tool for wide-field imaging and structured illumination applications. Recently, techniques have been developed to reconstruct high-resolution, wide-field images from multiple Fourier-space measurements~\cite{zheng2013wide}. In addition to imaging, Fourier-space techniques are also being used for optical trapping. For example, by leveraging the Fourier transform properties of an objective, researchers have shown how spatial light modulators can be used to simultaneously trap many particles in arbitrary 3D configurations~\cite{curtis2002dynamic,liang2014Spatial}.

Adapting optical microscopes to Fourier-space imaging and manipulation can introduce unexpected challenges. For example, even the simple task of focusing in Fourier space can lead to counterintuitive results. Whereas real-space alignment can be readily achieved by focusing on fine features in the image, such fine features in Fourier space are generally the product of aberrations (e.g., distortion near the pupil edge). In this context, Fourier microscopy raises a number of design choices that are distinct from real-space imaging. Specifically, most commercially available objectives and configurations have been optimized for different real-space applications. Yet, it is unclear which of these corrections are most important for quantitative Fourier imaging and also to what extent the optics for Fourier imaging will introduce additional aberrations. 

The purpose of this paper is to systematically evaluate the optical design choices and experimental parameters inherent in Fourier microscopy, including which microscope objectives are best suited for Fourier imaging, the ideal collection method and the relative advantages of different Bertrand lens configurations. To quantitatively examine these design choices, we first model complete microscope systems in Zemax. Detailed information about commercial microscope objectives and tube lenses are inferred by examination of published patents from Nikon, Olympus, and Zeiss. (As a potential resource to readers, we provide Zemax files for all optical elements as well as the combined systems in Ref.~\cite{kurvits2015Figshare}.) Based on ray-tracing analysis, we show that the ideal objective for Fourier microscopy is a plan-corrected apochromat with a high numerical aperture and low magnification. Furthermore, we show that placing the Bertrand lens into the ``infinity" space between the objective and tube lens can yield significant imaging improvements.


\section{Introduction to Fourier Microscopy}

Fourier microscopy typically involves three basic components: a microscope objective, a tube lens, and a Bertrand lens. This paper focuses on the application of Fourier imaging to modern microscopes, where an infinity-corrected microscope objective is used in combination with a tube lens to produce an image of the object at the exit port of the microscope. The addition of a Bertrand lens allows one to image the Fourier transform of the object by effectively re-imagining the objective's BFP; in this way, the tube lens and Bertrand lens can be seen as a two-lens optical system used to image the objective's BFP. For the purpose of this paper, we assume that the Bertrand lens will be a standard achromatic doublet designed for visible wavelengths. The term Bertrand lens is generally used to describe a lens that performs a Fourier transform without changing the position of the conjugate plane. For simplicity here though, we refer to any lens used to perform a Fourier transform as a Bertrand lens.

As shown in Fig.~\ref{FourierConfigs}, there are four commonly used configurations for BFP imaging. The first two simply place an achromatic doublet (Thorlabs AC254-050-A and AC254-100-A, respectively) either before~\cite{taminiau2012quantifying,schuller2013orientation,karaveli2013time} or after~\cite{hartmann2013radiation,dodson2014wide} the microscope's image plane as shown in Fig.~\ref{FourierConfigs}(a) and Fig.~\ref{FourierConfigs}(b), respectively. The first configuration typically limits the Bertrand lens to have a focal length of $\leq$50 mm (due to finite accessible space before the image plane) and is therefore limited in its magnification of the BFP. However, this configuration allows for simple switching between Fourier and real-space imaging by inserting or removing the Bertrand lens. The second configuration allows for greater magnification of the BFP image, but cannot be used for real-space imaging without additional optics.

The remaining two configurations attempt to overcome these limitations by using relay optics to move the image plane further from the microscope exit port. Figure~\ref{FourierConfigs}(c) shows a modified 4f relay (Thorlabs AC254-100-A and AC254-050-A), where the second of the two lenses can be replaced by one with twice the focal length to obtain a real-space image~\cite{shegai2011angular,shegai2011bimetallic}. The alternate configuration in Fig.~\ref{FourierConfigs}(d) uses a pair of relaying lenses (Thorlabs AC254-200-A and AC254-150-A) followed by an achromatic pair (Thorlabs MAP104040-A) that can be removed in order to obtain a real-space image~\cite{curto2013multipolar,hancu2013multipolar,curto2015personal}. Although these two designs offer greater flexibility, the additional elements significantly increase alignment difficulty and tolerancing errors. In the following analysis, we will weigh the advantages and disadvantages of these common configurations while also identifying a new approach.

\section{Optical Modeling of Commercial Microscopes}

Before considering the imaging configurations, it is possible to determine the ideal microscope objective. While microscope manufacturers do not tend to supply consumers with detailed models, their patent applications for objective lenses often specify a great deal of information. These patents often included the radius of curvature, thickness, refractive index, and Abbe number for each optical surface in the objective and associated tube lens. Therefore, by searching the patent literature, it is possible to infer likely designs for commonly used objectives.  Although these patents do not include details about any surface coatings used to minimize reflections (and therefore cannot be used to model overall system throughput), they do specify enough information for modeling optical aberrations. 

In this section, we provide the rationale by which we examined the patent literature and came to identify likely designs for commercially available components. To guide our search, we have sought to identify patent applications that were submitted near the commercial release date of new objectives, e.g. identifying a 2004 patent application from Nikon ~\cite{mandai2006Immersion} describing TIRF objectives with "NA larger than 1.45" that predates the 2005 release of their 1.49 NA TIRF objectives~\cite{nikon2005release}. Wherever possible, we have then tried to use physical examination (i.e., disassembly of objectives and tube lenses) to confirm basic design properties such as the number of elements and their relative curvatures. For every objective examined, we did notice slight discrepancies in specific details, e.g. in the curvatures or thicknesses of some lenses, but discrepancies are to be expected from patent specifications. Despite being sold under the same commercial name, the designs for objectives and tube lenses are presumably subject to continuous improvements and modifications. For example, over the last 20 years, the removal of environmentally hazardous materials from optical glasses (e.g., arsenic, lead and HCFCs) has required the redesign of many optical components~\cite{nikon2015Environmental}. Nevertheless, we believe these detailed patents are still a helpful source for design optimization, particularly for understanding the differences between objective classes, even if the final objectives may vary from these exact specifications.

\subsection{Glass Determination}
Commercial patents generally specify optical materials by their Abbe number and refractive index at the d-line (587.5 nm). With these parameters alone, optical models are only accurate over a relatively small range of wavelengths near the d-line. To enable more accurate analysis over the full visible spectrum, we performed a glass substitution using Zemax's built in glass catalogs. Where possible, we replaced each material with tabulated data from a ``standard'' or ``preferred'' glass with an Abbe number and index identical to the patent specification. If an exact match could not be found for the Abbe number, we looked for the closest match that still provided an exact index match. Where no index match was possible, we simply used the original Abbe number and refractive index. These substitutions greatly improved chromatic aberrations and brought them qualitatively in line with the spectrally dependent field curvature plots provided in the patents. We further validated this process by examining the glass manufacturers. For the Zeiss objective, it was possible to use Schott glass for all but one surface. (Carl Zeiss Microscopy and Schott are both subsidiaries of the Carl Zeiss Foundation.) Similarly for Olympus, all but one surface was Ohara glass. (Ohara lists Olympus as one of its major customers~\cite{ohara2015Corp}.) Finally, Nikon objectives appear to use Schott and a variety of Japanese optical glass manufacturers, including its subsidiary Hikari.

\subsection{Tube Lens}
The first element we were able to identify in commercial microscopes is the tube lens, because it is a relatively simple component. Since tube lenses are also integral to all infinity-corrected microscope designs, they are often defined in multiple patents. By searching the patent literature, we found numerous patents from Nikon~\cite{mandai2006Immersion,watanabe1998Immersion,furutake1999Immersion,kudo1999Immersion,furutake2000Microscope,yamaguchi2003Immersion,okuyama2003Liquid,watanabe2004Liquid,yamaguchi2007immersion,yoshida2011Immersion,yamaguchi2012Immersion}, Olympus~\cite{suzuki1996Immersion,suzuki1996ImmersionTube,suzuki1997Immersion,fujimoto2007Immersion,konishi2009Immersion,kasahara2013Immersion1,kasahara2013Immersion} and Zeiss~\cite{wartmann2006Liquid,muchel2007Tube,shi2010Immersion} in which the same tube lens was specified for a given manufacturer. In addition, we were able to obtain and disassemble lenses from all three manufacturers. We found that the Zeiss tube lens (Part number: 452960) matched the single element design described in objective patents. We also found that the Nikon tube lens for Ti-U inverted microscope (Part Number: MEA53210-1XL) agrees with the patents in the number of components and their relative curvatures, but component thicknesses differed by $\sim$1~mm. However, the Olympus tube lens (Part Number: U-TLU-1-2) had a different number of elements than the patent specification. Despite these discrepancies, we are confident in the basic designs as they have been listed in many patents spanning several decades and, therefore, seem to be the tube lenses used by the manufacturers to specify objective performance. In Tables~\ref{nikonTubeLens} -~\ref{zeissTubeLens} of Appendix A, we present the tube lens implementations specified by Nikon, Olympus, and Zeiss patents.


\subsection{Objectives}

A number of different objectives were chosen from Nikon, Olympus, and Zeiss based on their availability in the patent literature and in order to cover a large range of NA and aberration corrections. A detailed list of objective manufacturers, magnifications, NA, aberration corrections and their associated patents can be found in Table~\ref{table:Objectives}. Although we cannot be certain that these are exact matches to commercial products, we still believe that the qualitative results obtained from analyzing these patents can help identify the optimal objective class for Fourier imaging.

\begin{table}[b]
\caption{Patent Sources used for Objective Modeling}
\begin{tabular}{c c c l c}
\hline
\underline{\smash{U.S. Patent \# [Ref. ]}} &
\underline{\smash{Assignee}} & \underline{\smash{Mag.}} & \underline{\smash{NA}} & \underline{\smash{Class/Correction}}\\
5,517,360~\cite{suzuki1996Immersion} & Olympus & 60 & 1.4 & Plan Apo\\
5,659,425~\cite{suzuki1997Immersion}& Olympus & 100 & 1.65 &  high index Apo\\
6,504,653~\cite{matthae2003High} & Zeiss & 100 & 1.45 & TIRF \\
6,519,092~\cite{yamaguchi2003Immersion} & Nikon & 60 & 1.4 & Plan Apo \\ 
& Nikon & 100 & 1.4 & Plan Apo \\
7,046,451~\cite{mandai2006Immersion} & Nikon & 60 & 1.5$^*$  & TIRF \\
 & Nikon & 100 & 1.5$^*$ & TIRF \\
7,889,433~\cite{yoshida2011Immersion} & Nikon & 60 & 1.25 &  water immersion \\\hline
\end{tabular}
$\quad^*$~modeled as 1.49~NA to match possible commercial realization
\label{table:Objectives}
\end{table}

To ensure that the Zemax models faithfully represent the patent specifications, we first reproduced the d-line distortion and spherical aberration figures in the patents. As discussed above, we then performed a detailed glass substitution process; this allowed us to confirm the wavelength dependent spherical aberration plots. Note that the glass substitution procedure does not directly affect our analysis, because we perform all subsequent aberration calculations at the d-line. Nevertheless, we chose to perform the glass substitutions to ensure we correctly modeled the objectives and to create a library of objectives that would be of broader use. 

In addition to limited information on the optical glasses that are used, almost none of the patents specify component semi-diameters (which are relevant to vignetting). It is not possible with these patent specifications to assume that the objectives could achieve their full NA over the full field of view, because this would require physically impossible lens surfaces, e.g. lenses that intersect other lenses and/or themselves. In order to determine the semi-diameters, we allowed these values to float such that the objective collects all the light from an on-axis point source which emits at the specified NA. Using this process, and comparing the resulting semi-diameters to patents where semi-diameters were specified~\cite{suzuki1996Immersion,matthae2003High}, we were able to validate this approach. Specifically, for Ref.~\cite{suzuki1996Immersion}, there was $<2$ $\mu$m error on all surfaces. For Ref.~\cite{matthae2003High}, there was $<0.5$ mm error on all surfaces and an exact match on the surface that is the primary source of vignetting. For the results of the glass substitution and semi-diameter determination, see Tables~\ref{nikon1p4_60x} -~\ref{zeiss1p45_100x} in Appendix B. Figure ~\ref{objectiveSchematic} shows an example of the resulting modeled surfaces from a Nikon 1.4NA, 60x objective.

\begin{figure}
\includegraphics[width=\columnwidth]{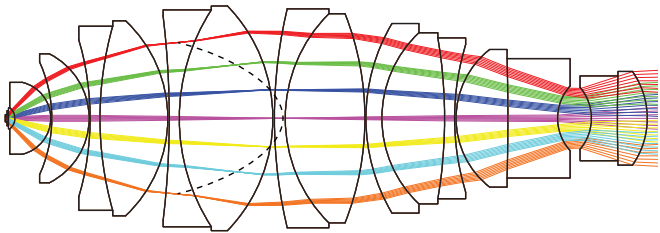}
\caption{Schematic of 60x Objective from Nikon patent~\cite{yamaguchi2003Immersion}. The full specification, included in Appendix B, contains 23 surfaces corresponding to eight component groups composed of eight glass types. Colored lines represent rays emitted at various angles from object plane. Dashed black line shows surface where these rays cross and form a Fourier "plane" inside the microscope objective. Note that this Fourier "back-focal-plane" is a highly curved surface in all the objectives examined here.}
\label{objectiveSchematic}
\end{figure}

\section{Comparative Analysis of Microscope Objectives}
For real-space imaging of structures and specimens, a large variety of microscope objectives have been developed, each with different optimization metrics in mind. These includes apochromatic (low chromatic aberrations), Plan (flat field and low distortion), and TIRF (small depth of field) objectives. However, since these were all designed for real-space imaging, their performance for Fourier imaging is not well known. Here, we examine the effect of aberration corrections, nominal magnification, and numerical aperture as well as the choice of confocal or full-field collection on the quality of the resulting Fourier image.

\begin{figure}[t!]
\includegraphics{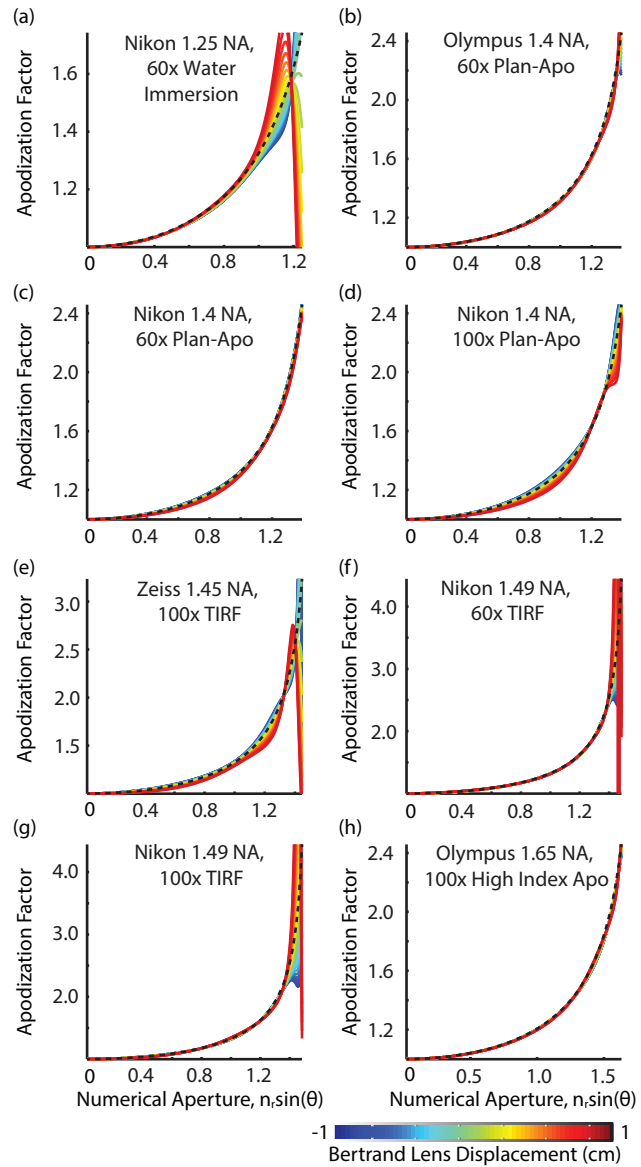}
\caption{Apodization factor as a function of BL position in the configuration shown in Fig.~\ref{FourierConfigs}(a). Significant deviations from the ideal $1/\cos(\theta)$ apodization factor (dashed black line) are seen for TIRF (e-g) and high NA water immersion objectives (a), whereas Apo (h) and Plan-Apo (b-d) objectives perform well even over a relatively large BL displacement of $\pm1$ cm.}
\label{apodization}
\end{figure}

\subsection{Objective Class and Alignment Sensitivity}
When selecting an objective for Fourier-space imaging, it is well known that a great deal of information exists beyond the critical angle for total internal reflection (e.g., $NA=1$ for a sample-air interface). Thus, a common approach is to maximize angular collection by using the largest possible NA objective. However, image distortion increases significantly with NA. This is especially true when the NA approaches the refractive index of the immersion medium, $n_{r,imm}$. Indeed, a key measure in the design of high NA objectives is the ratio ${NA}/n_{r,imm}$. TIRF patents often present their claims in terms of such ratios; for example, Ref.~\cite{matthae2003High} specifies a TIRF objective relative to a $NA/n_{r,imm}>0.938$ condition. As this number approaches unity and higher angle rays are collected, both aberrations and tolerancing errors are expected to increase. 

To examine this effect, we plot in Fig.~\ref{apodization} the apodization factors for the objectives listed in Table~\ref{table:Objectives} as a function of defocusing a Bertrand lens (50 mm achromatic doublet) placed before the image plane. For each objective, distortion data was extracted as a function of Bertrand lens position and apodization factors were calculated according to Ref.~\cite{sheppard2007imaging}. The apodization factor is a dimensionless scaling term that describes how projected ray density increases with increasing field angle ($\theta$). This factor is derived from the distortion introduced by the optical system, but is generally assumed to be $1/\cos(\theta)$ for systems obeying the Abbe sine condition (i.e., systems with a wide field of view such as microscopes)~\cite{lieb2004single}. Note that the apodization factor was chosen as our metric of objective quality, because it is typically the only correction made to Fourier-space images to take into account the optical setup~\cite{lieb2004single,lee2011planar,bharadwaj2011electrical}. We specifically examine defocus misalignment as it is often the most difficult tolerancing error to correct, because it tends to couple with distortion in common Fourier microscopy configurations. While decentering and tilt of the Bertrand lens also lead to undesirable changes to the Fourier image, they are relatively easy to observe and correct for even in Fourier space. (For example, by observing fluorescence from an isotropic thin film emitter, tilt and decenter are seen as clear radial asymmetries in the resulting Fourier image, which are readily corrected.) Finally, we chose to use the BIP Fourier imaging configuration shown in Fig.~\ref{FourierConfigs}(a) due to its simple design, which allowed for easy determination of the objective's influence. 

To analyze the results in Fig.~\ref{apodization}, consider first the objectives that use a standard $\sim$1.515 refractive index immersion oil. As can be seen from a comparison to the ideal $1/\cos(\theta)$ apodization shown as a dashed black line, the Plan-Apo objectives in Fig.~\ref{apodization}(b-d) perform well over the full numerical aperture, whereas TIRF objectives in Fig.~\ref{apodization}(e-g) perform quite poorly at NA values beyond 1.3. Note that the sharp peaks and sudden drop off at high NA values are features commonly seen during experimental alignment of the Bertrand lens. Interestingly, the 1.65 NA objective performs quite well, but this is explained by the fact that it uses a high $\sim$1.780 refractive index immersion oil. The high index 1.65 NA objective thus has a relatively low $NA/n_{r,imm}$ ratio of 0.927, which is comparable to that of the 1.4 NA objectives. In contrast, the 1.25 NA water immersion objective, which has a relatively high $NA/n_{r,imm}$ ratio of $\sim$0.938, performs quite poorly. This underscores the point that NA alone may not be the best criteria in selecting an objective for Fourier microscopy. In particular, one should be careful about the use of TIRF and other high $NA/n_{r,imm}$ objectives, especially in applications where a high NA Plan-Apo objective may be suitable. 

For completeness, we note that the objective performance trends discussed above for the BIP configuration were also observed for the AIP configuration in Fig. 1(b) as well as the preferred configuration discussed in Section 5 where the Bertrand lens is placed before the tube lens. While the choice of imaging configuration does impact the magnitude of tolerancing errors, the Plan-Apo and high-index Apo objectives consistently demonstrated lower tolerancing errors than the TIRF and 1.25 NA water immersion objectives.

\subsection{Magnification and Fourier Space Resolution}
\begin{figure}
\includegraphics[width=\columnwidth]{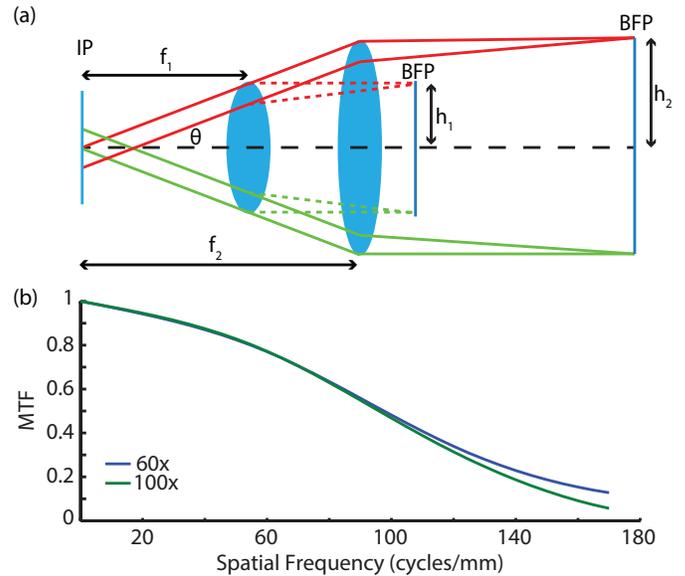}
\caption{Effect of magnification on Fourier image size and resolution.  (a) Schematic demonstrating how, for a given NA, the objective's BFP is larger for smaller nominal magnifications. Where f is the effective focal length, d is the semi-diameter of the BFP image, and $\theta$ is the half angle determined by the NA of the objective. (b) Comparison of modulation transfer functions for the imaging configuration shown in Fig.~\ref{FourierConfigs}(a) using 60x and 100x Nikon 1.4 NA Plan-Apo objectives.}
\label{MTFCompare}
\end{figure}

In addition to aberration corrections and NA, magnification is an important parameter for image quality and resolution. For real-space imaging, the choice is fairly obvious, the highest NA and magnification will allow you to resolve the smallest features. However, for Fourier-space imaging, something counterintuitive occurs. For the same NA, you actually obtain a larger Fourier image when using an objective with a lower nominal magnification. To see why, consider the simple schematic shown in Fig.~\ref{MTFCompare}, which shows how the Fourier image is formed at the BFP of a single lens. As the focal length is increased (in order to decrease the magnification for a fixed tube lens), the semi-diameter must increase in order to maintain the same NA. Thus, as can be seen in Eq.~\ref{Equation:Magnification} below, the BFP semi-diameter height, $h$, is inversely proportional to the objective's nominal real-space magnification, $M$,  specified by the manufacturer:
\begin{equation}
h=\frac{f_t}{M}\left(\left(\frac{n_r}{NA}\right)^2-1\right)^{-1/2},
\label{Equation:Magnification}
\end{equation}

\noindent where $f_t$ is the focal length of the tube lens, $n_r$ is the immersion oil index of refraction, and NA is the numerical aperture of the objective. However, there's still a question as to whether this improves the angular resolution; the image may be larger, but it may also be more blurred out. To demonstrate that this is not the case, we examine the modulation transfer function (MTF) for two comparable systems with differing magnification. (The MTF is defined as the contrast ratio that would be observed at the image plane for a sine wave object of a given spatial frequency.) Specifically, we plot the MTF for two Nikon objectives of the same NA but different magnifications (objectives 3 and 4 in Table 1). Note that, not only do these two objectives have the same NA, their specifications came from the same patent~\cite{yamaguchi2003Immersion} making them the ideal comparison. As can be seen in Fig.~\ref{MTFCompare}(b), both objectives have a very similar MTF (and therefore comparable angular resolutions). Thus, the lower magnification objective maintains BFP image quality while increasing its size. (However, it is important to note that the smaller BFP images from high magnification objectives may still be desirable when working with low light samples, e.g. weak single emitters.)


\subsection{Vignetting Reductions of Effective NA}

Although these objectives were all designed for low distortion wide-field imaging, they still suffer from reduced throughput at the edges of the field of view. For real-space imaging, this simply leads to a darker image at the edges than in the center. However, for Fourier imaging, vignetting decreases the observed intensity (collection efficiency) at large angles when imaging the full field of view. This, in turn, can lead to quantitatively different results when fitting the resulting radiation pattern. Thus, it is important to consider the effect of the collection area (which is often linked to the excitation spot size).

\begin{figure}
\centering\includegraphics[width=\columnwidth]{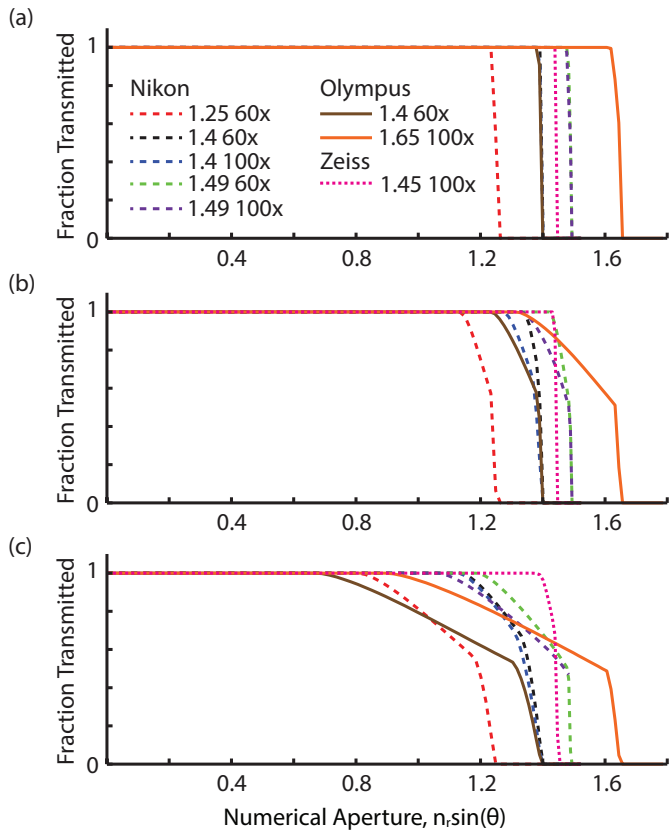}
\caption{Fraction of light transmitted for (a) 10 $\mu$m spot, (b) 100 $\mu$m spot and (c) full field of view for the objectives listed in Table.~\ref{table:Objectives} using the imaging configuration shown in Fig.~\ref{FourierConfigs}(a).}
\label{Vignetting}
\end{figure}

For collection from a 1~$\mu$m confocal region at the center of the field of view, vignetting is negligible, and all objectives transmit essentially 100\% of rays up to their full NA. However, for wide-field collection, almost all objectives experience serious vignetting far short of their designed NA. To illustrate this, we plot the fraction of transmitted rays in Fig.~\ref{Vignetting} for three different collection areas, ranging from a 10~$\mu$m diameter area in Fig.~\ref{Vignetting}(a) to a 100~$\mu$m diameter area in Fig.~\ref{Vignetting}(b) to the specified full-field in Fig.~\ref{Vignetting}(c). (Note that the full-field size depends on the manufacturer and magnification, because fields are generally specified by image, rather than object, size. Nikon and Olympus specify the full-field to be a 22~mm image, whereas Zeiss specifies a 25~mm image). While the vignetting effects within a 10~$\mu$m central area are minor, it is clear from Fig.~\ref{Vignetting} that increasing the field beyond 100~$\mu$m (or operating farther than 50~$\mu$m from the center of the field) will reduce the effective NA and can distort the Fourier image. Thus, although wide-field excitation of a fluorescent sample can greatly improve the signal-to-noise ratio, it will lead to an undesirable quantitative modification to the Fourier image in almost all microscope objectives and should therefore be avoided.


\section{Comparative Analysis of Imaging Configurations}
\begin{figure}
\centering\includegraphics[width=\columnwidth]{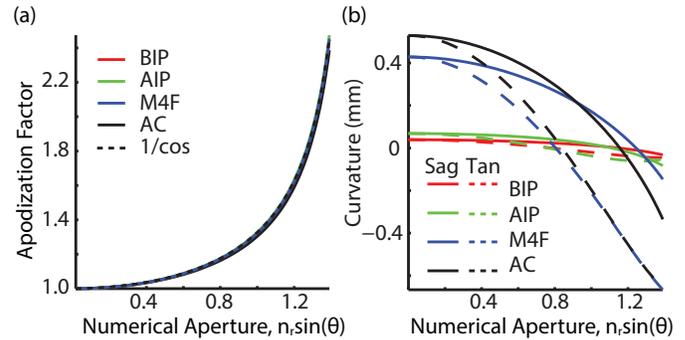}
\caption{Comparison of different Fourier imaging configurations with a Nikon 1.4NA 60x objective. BIP, AIP, M4F and AC correspond to the configurations shown in Fig.~\ref{FourierConfigs}(a-d), respectively. (a) Apodization factor for each configuration shown together with the ideal $1/\cos(\theta)$ apodization factor (dashed black line) for a system obeying the Abbe sine condition. (b) Sagittal and tangential field curvature as a function of field angle for four configurations shown in Fig.~\ref{FourierConfigs}.}
\label{configurationcompare}
\end{figure}

From the analysis in Section 4, we have determined that the best choice for Fourier-space imaging is a plan corrected apochromatic objective with a low magnification and a high NA. (Although all calculations were done at the d-line and therefore chromatic aberrations were not considered, the specific objectives that performed the best in our modeling were Plan-Apos. We would also recommend the use of apochormatic objectives in general, because the majority of fluorescence experiments involve broadband emitters.) With this in mind, we will now consider the relative performance of the four imaging configurations shown in Fig.~\ref{FourierConfigs}. To compare these four configurations, we use the 60x 1.4NA Nikon Plan-Apo objective, and extract the field curvature and apodization factor at the detector plane. (As mentioned above, the apodization factor is often the only modification used to account for the optical setup in the literature.) As can be seen in Fig.~\ref{configurationcompare}(a), the apodization factors are nearly identical for all four configurations. However, Fig.~\ref{configurationcompare}(b) shows clear differences in the field curvatures. The field curvature is plotted in units of mm as the distance between the paraxial image plane and the real image plane as a function of $n_r\sin(\theta)$. While the AIP and BIP configurations have field curvatures on the order of 0.1~mm, the M4F and AC relay configurations exhibit curvature on the order of 1~mm. Therefore, it seems that the simpler AIP and BIP configurations may be slightly better, but given the small curvature (i.e. $\leq$1.3~mm) for all systems, it is not clear that this would lead to a significant imaging improvement.

    The choice between the four Fourier imaging configurations in Fig. ~\ref{FourierConfigs} is often limited by practical constraints. For example, it is often useful to place additional optics (e.g., polarizers, beamsplitters, filters, etc.) in the light path. Making the focal length of the Bertrand lens as long as possible can also be desirable, because it makes the BFP image on the detector larger and therefore reduces pixelation. Thus, despite being the simplest modification to a standard microscope, the BIP configuration in Fig.~\ref{FourierConfigs} is overly restrictive. Requiring the detector to be at the exit port image plane puts significant limits on the focal length of the Bertrand lens and leaves little room for other optical components. The AIP configuration alleviates both of these issues, but cannot be used to obtain real-space images at the same detector plane. Finally, the M4F and AC relay configurations are the most flexible in magnification and allow for the removal or replacement of lenses to obtain real-space images. However, this increased flexibility comes at the cost of increased field curvature as well as increased complexity and associated tolerancing errors. 

\begin{figure}[t]
\includegraphics[width=\columnwidth]{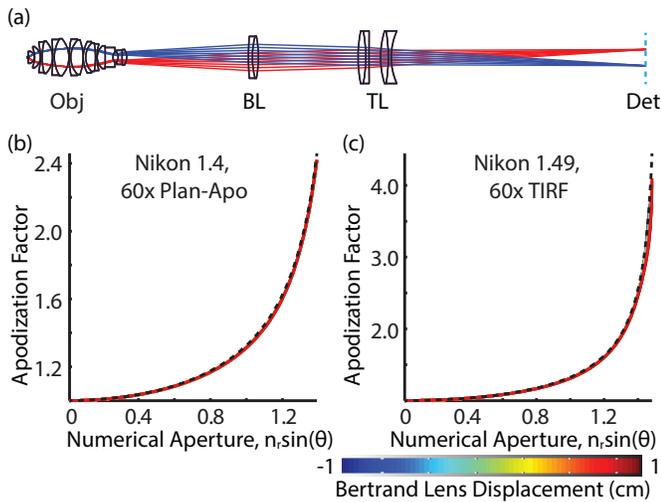}
\caption{Schematic of Fourier microscopy configuration where a Bertrand lens is placed before the tube lens (a) along with apodization factor as a function of Bertrand lens position for a Nikon 1.4 NA, 60x Plan-Apo objective (b) and a Nikon 1.49 NA, 60x TIRF objective. A significant improvement in tolerancing errors can be seen by comparison to Fig.~\ref{apodization}(c,f).}
\label{bertrandBBL}
\end{figure}

    Interestingly, there is one other place where we have access to the beam path in an standard inverted microscope, namely in the ``infinity" space between the objective and tube lens. Although this space is typically occupied by excitation and emission filters, these are easily removed, and Fourier imaging optics can be put in their place. The advantage of this setup is that the Bertrand lens, which now effectively acts as a tube lens, can be placed such that it directly focuses on the objective's BFP, as shown in Fig.~\ref{bertrandBBL}(a). This configuration combines the simplicity of the one-lens configurations in Fig.~\ref{FourierConfigs}(a,b) and the BFP magnification flexibility of the relay configurations in Fig.~\ref{FourierConfigs}(c,d) without curvature or tolerancing complications. As can be seen by comparison of Fig.~\ref{bertrandBBL}(b,c) to Fig.~\ref{apodization}(c,f), tolerancing errors are greatly reduced when the Bertrand lens is placed before the tube lens. This is primarily due to the fact that the Bertrand lens is now directly imaging the back-focal-plane of the objective, and therefore forms an infinite conjugate pair with the tube lens, whereas in all other configurations they form a finite conjugate pair.
    
    Importantly, the configuration shown in Fig.~\ref{bertrandBBL}(a) also provides a major advantage for system alignment, because deviations in the position of the Bertrand lens cause almost pure defocus. In the four common configurations shown in Fig.~\ref{FourierConfigs}, movement of the Bertrand lens leads to both defocus and distortion, and it is distortion changes near the pupil edge that often produce the sharpest features during BFP alignment. However, when positioning the Bertrand lens in infinity space, the lack of distortion effects means that sharp features in the resulting Fourier image do correspond to good focus and, thus, proper alignment.


\section{Conclusion}

In conclusion, we have used patent information to model a range of microscope objectives in order to determine the ideal Fourier-space imaging setup. We have shown that, despite the significantly larger angles available to TIRF objectives, large deviations from ideal apodization can lead to severely aberrated Fourier images, especially for commonly used imaging configurations. It is therefore best to use a high NA, Plan-Apo objective with a low magnification. Also, despite the gains in signal-to-noise when collecting from the full field of view, vignetting leads to an undesirable decrease in throughput at high angles and should therefore be avoided by using on-axis small-area collection. Finally, when using a standard commercially available microscope, the choice of imaging configuration outside the microscope is primarily a practical concern. However, with simple modifications, placing a lens between the objective and tube lens can significantly improve imaging performance and simplify the alignment process. We hope that these results, together with the tabulated surface data in the Appendices and Zemax models in Ref.~\cite{kurvits2015Figshare}, will help advance the use of Fourier microscopy in nano-optics and nanophotonics. We also hope the methods and analysis presented here will be of interest to other fields, such as biological imaging, where complete models of microscope objectives and systems could help optimize high resolution techniques.

\vspace{\baselineskip}
{\setlength{\parindent}{0cm}
\textbf{Funding.} Air Force Office of Scientific Research (MURI FA9550-12-1-0488); US Department of Education (GAANN Award P200A090076)

\vspace{\baselineskip}
\textbf{Acknowledgment.} The authors thank Alberto G. Curto, Christopher M. Dodson, and Jason P. McClure for helpful discussions.}

\appendix
\setlength{\abovecaptionskip}{0pt}
\setlength{\belowcaptionskip}{10pt}

\section{Tube Lens Surface Specifications}\label{appendixA}
Listed below are the specifications for the most likely implementations of Nikon, Olympus, and Zeiss tube lenses inferred from the patent literature, where r is the radius of curvature, d is the thickness, h is the semi-diameter height, and Mfr is the glass manufacturer. (Semi-diameters are set to the smallest clear aperture measured from actual tube lenses.)
\topcaption{Tube Lens from Nikon Patents ~\cite{mandai2006Immersion,watanabe1998Immersion,furutake1999Immersion,kudo1999Immersion,furutake2000Microscope,yamaguchi2003Immersion,okuyama2003Liquid,watanabe2004Liquid,yamaguchi2007immersion,yoshida2011Immersion,yamaguchi2012Immersion}}
\label{nikonTubeLens}
\begin{supertabular}{ c c c c c c }
\hline
\underline{\smash{Surf       }}&\underline{\smash{r(mm)}}&\underline{\smash{d(mm)}}&\underline{\smash{h(mm)}}&\underline{\smash{Glass}}&\underline{\smash{Mfr}}\\
1&75.043&5.1&15.9385&E-SK10&Hikari\\
2&-75.043&2&15.9385&J-LAF7&Hikari\\
3&1600.58&7.5&15.9385&&\\
4&50.256&5.1&15.9385&BASF6&Schott\\
5&-84.541&1.8&15.9385&KZFH1&Hikari\\
6&36.911&168.4117&15.9385&&\\\hline
\end{supertabular}

\topcaption{Tube Lens from Olympus Patents~\cite{suzuki1996Immersion,suzuki1996ImmersionTube,suzuki1997Immersion,fujimoto2007Immersion,konishi2009Immersion,kasahara2013Immersion1,kasahara2013Immersion}}
\label{olympusTubeLens}
\begin{supertabular}{ c c c c c c }
\hline
\underline{\smash{Surf       }}&\underline{\smash{r(mm)}}&\underline{\smash{d(mm)}}&\underline{\smash{h(mm)}}&\underline{\smash{Glass}}&\underline{\smash{Mfr}}\\
1&68.7541&7.7321&13.97&S-FSL5&Ohara\\
2&-37.5679&3.4742&13.97&H-ZLAF52&CDGM\\
3&-102.8477&0.6973&13.97&&\\
4&84.3099&6.0238&13.97&S-LAH60&Ohara\\
5&-50.71&3.0298&13.97&BPH35&Ohara\\
6&40.6619&156.9522&13.97&&\\\hline
\end{supertabular}

\topcaption{Tube Lens from Zeiss Patents~\cite{wartmann2006Liquid,muchel2007Tube,shi2010Immersion}}
\tablehead{}
\tabletail{}
\label{zeissTubeLens}
\begin{supertabular}{ c c c c c c }
\hline
\underline{\smash{Surf       }}&\underline{\smash{r(mm)}}&\underline{\smash{d(mm)}}&\underline{\smash{h(mm)}}&\underline{\smash{Glass}}&\underline{\smash{Mfr}}\\
1&189.417&10.9&12.5&N-BALF4&Schott\\
2&-189.417&160.7711&12.5&&\\\hline
\end{supertabular}

\section{Objective Surface Specifications}\label{appendixB}
Listed below are the specifications for the microscope objectives listed in Table~\ref{table:Objectives}, where the notation is the same as described above in Appendix A. Note that for surfaces where an exact glass match could not be obtained (and for the immersion oil for every objective), the index and Abbe number are given at the d-line instead.
\topcaption{60x, 1.4NA Objective from Nikon Patent~\cite{yamaguchi2003Immersion}}
\label{nikon1p4_60x}
\begin{supertabular}{ c c c c c c }
\hline
\underline{\smash{Surf       }}&\underline{\smash{r(mm)}}&\underline{\smash{d(mm)}}&\underline{\smash{h(mm)}}&\underline{\smash{Glass}}&\underline{\smash{Mfr}}\\
1&Infinity&0.17&0.183&1.52216, 58.80&\\
2&Infinity&0.15&0.39834&1.51536, 41.36&\\
3&Infinity&0.65&0.76209&S-NSL3&Ohara\\
4&-1.332&3.6&1.0598&LASF35&Schott\\
5&-3.716&0.1&3.6153&&\\
6&-13.716&3.75&5.6585&GFK70&Sumita\\
7&-7.247&0.1&6.4791&&\\
8&-27.891&1&7.8796&J-F5&Hikari\\
9&34.23&6.8&9.2544&GFK70&Sumita\\
10&-13.453&0.15&9.7985&&\\
11&-84.754&1&10.2849&J-KZFH1&Hikari\\
12&20.048&9.4&10.8992&LITHO-CAF2&Schott\\
13&-16.266&0.15&11.288&&\\
14&47.671&1.1&11.0093&J-KZFH1&Hikari\\
15&14.802&8&10.5143&LITHO-CAF2&Schott\\
16&-28.664&0.1&10.4895&&\\
17&18.671&1.6&9.5306&J-KZFH1&Hikari\\
18&11.816&6.3&8.6046&LITHO-CAF2&Schott\\
19&-48.478&1&8.0904&1.526820, 51.35&\\
20&25.246&0.15&7.4167&&\\
21&8.784&5.2&6.9181&GFK70&Sumita\\
22&-238.404&5&6.0015&S-LAH63&Ohara\\
23&4.823&3.4&3.2407&&\\
24&-4.801&2.6&3.1155&S-LAH63&Ohara\\
25&204.674&3&4.2684&FD60-W&Hoya\\
26&-8.172&&4.7147&&\\\hline
\end{supertabular}

\topcaption{100x, 1.4NA Objective from Nikon Patent~\cite{yamaguchi2003Immersion}}
\label{nikon1p4_100x}
\begin{supertabular}{ c c c c c c }
\hline
\underline{\smash{Surf       }}&\underline{\smash{r(mm)}}&\underline{\smash{d(mm)}}&\underline{\smash{h(mm)}}&\underline{\smash{Glass}}&\underline{\smash{Mfr}}\\
1&Infinity&0.17&0.11&1.52216, 58.80&\\
2&Infinity&0.15&0.39664&1.51536, 41.36&\\
3&Infinity&0.6&0.75712&S-NSL3&Ohara\\
4&-1.113&3.3&0.94294&LASF35&Schott\\
5&-3.32&0.1&3.2671&&\\
6&-12.476&3.261&5.1827&J-PSK03&Hikari\\
7&-6.818&0.15&5.9356&&\\
8&-28.872&1&7.1835&1.52682, 51.35&\\
9&20.752&7.77&8.6544&J-FKH1&Hikari\\
10&-12.157&0.2&9.2943&&\\
11&-151.459&1&9.7384&F5&Schott\\
12&18.68&8.227&10.1252&LITHO-CAF2&Schott\\
13&-16.862&0.2&10.4318&&\\
14&25.434&1&10.0854&J-KZFH1&Hikari\\
15&11.981&7.597&9.4033&LITHO-CAF2&Schott\\
16&-28.918&0.2&9.2799&&\\
17&13.722&1.5&8.04&LAF7&Hoya\\
18&9.019&5.74&7.0106&LITHO-CAF2&Schott\\
19&-24.314&1.5&6.5153&KZFH2&Hikari\\
20&20.929&13.659&5.7806&&\\
21&-115.034&1&3.3347&M-TAF1&Hoya\\
22&7.657&3&3.1723&H-ZF7L&NHG\\
23&-7.822&1&3.0012&M-TAF1&Hoya\\
24&11.351&&2.7873&&\\\hline
\end{supertabular}

\topcaption{60x, 1.25NA Objective from Nikon Patent~\cite{yoshida2011Immersion}}
\label{nikon1p25_60x}
\begin{supertabular}{ c c c c c c }
\hline
\underline{\smash{Surf       }}&\underline{\smash{r(mm)}}&\underline{\smash{d(mm)}}&\underline{\smash{h(mm)}}&\underline{\smash{Glass}}&\underline{\smash{Mfr}}\\
1&Infinity&0.17&0.213&1.5244, 54.30&\\
2&Infinity&0.25&0.544&1.3326, 55.90&\\
3&-10&0.63&0.82735&LITHOSIL-Q&Schott\\
4&-1.051&2.82&0.94768&LAH55&Ohara\\
5&-2.921&0.1&2.8684&&\\
6&-12.431&2.75&4.5036&GFK68&Sumita\\
7&-6.681&0.15&5.2569&&\\
8&-63.897&1&6.4222&S-NSL36&Ohara\\
9&13.457&8.85&7.7123&E-FKH1&Hikari\\
10&-11.96&0.2&8.6884&&\\
11&-636.078&1&8.9946&LAH59&Ohara\\
12&17.16&9.05&9.1801&LITHO-CAF2&Schott\\
13&-13.417&0.2&9.748&&\\
14&17.111&1.2&9.3754&YGH51&Ohara\\
15&11.17&6.9&8.6363&LITHO-CAF2&Schott\\
16&-26.536&0.6&8.4528&&\\
17&27.985&1.1&7.3624&LAH59&Ohara\\
18&20.792&4.5&6.939&LITHO-CAF2&Schott\\
19&-13.585&1&6.3993&LAH59&Ohara\\
20&46.225&0.2&6.0693&&\\
21&7.409&5.9&5.828&E-FKH1&Hikari\\
22&-28.987&4.6&4.6985&BSM81&Ohara\\
23&3.708&2.9&2.545&&\\
24&-4.496&4.4&2.4988&J-PSK03&Hikari\\
25&36.446&3.7&3.6535&FL7&Hoya\\
26&-7.761&&4.1654&&\\\hline
\end{supertabular}

\topcaption{60x, 1.49NA Objective from Nikon Patent~\cite{mandai2006Immersion}}
\label{nikon1p49_60x}
\begin{supertabular}{ c c c c c c }
\hline
\underline{\smash{Surf       }}&\underline{\smash{r(mm)}}&\underline{\smash{d(mm)}}&\underline{\smash{h(mm)}}&\underline{\smash{Glass}}&\underline{\smash{Mfr}}\\
1&Infinity&0.17&0.2085&1.52210, 58.8000&\\
2&Infinity&0.13&0.81893&1.51299, 40.6812&\\
3&Infinity&0.75&1.5517&S-NSL3&Ohara\\
4&-2.243&3.85&1.6556&S-LAH79&Ohara\\
5&-3.827&0.1&3.827&&\\
6&-23.274&5&7.0632&GFK68&Sumita\\
7&-8.761&0.15&7.8999&&\\
8&-38.045&1&9.0961&E-F2&Hikari\\
9&16.326&11&10.819&GFK70&Sumita\\
10&-15.9&0.15&11.4868&&\\
11&331.735&1&11.4259&N-KZFS8&Schott\\
12&17&10.4&11.3175&LITHO-CAF2&Schott\\
13&-17.778&0.15&11.6131&&\\
14&34.108&1&10.8098&N-KZFS5&Schott\\
15&16.2&5.6&10.2122&LITHO-CAF2&Schott\\
16&-103.612&1&10.0258&&\\
17&17&4.1&9.0895&LITHO-CAF2&Schott\\
18&-129.879&1&8.6885&N-KZFS5&Schott\\
19&21.365&0.15&7.9004&&\\
20&9.002&6.1&7.3838&J-PSK03&Hikari\\
21&-48.082&2.65&6.3398&J-LASF015&Hikari\\
22&5.9&4.45&4.1233&&\\
23&-6.584&1&3.9245&S-LAH66&Ohara\\
24&20.8&3.4&4.5938&J-SF03&Hikari\\
25&-11.342&&4.9719&&\\\hline
\end{supertabular}

\topcaption{100x, 1.49NA Objective from Nikon Patent~\cite{mandai2006Immersion}}
\label{nikon1p49_100x}
\begin{supertabular}{ c c c c c c }
\hline
\underline{\smash{Surf       }}&\underline{\smash{r(mm)}}&\underline{\smash{d(mm)}}&\underline{\smash{h(mm)}}&\underline{\smash{Glass}}&\underline{\smash{Mfr}}\\
1&Infinity&0.17&0.125&1.52210, 58.8000&\\
2&Infinity&0.13&0.819&1.51299, 40.6812&\\
3&Infinity&0.93&1.5517&KF6&Schott\\
4&-2.322&3.95&1.8188&S-LAH79&Ohara\\
5&-3.939&0.15&3.9337&&\\
6&-38.362&4.3&7.7336&GFK68&Sumita\\
7&-9.799&0.1&8.1798&&\\
8&255.173&1&9.6314&N-KZFS5&Schott\\
9&16.559&7.5&10.3719&GFK70&Sumita\\
10&-20.805&0.15&10.5917&&\\
11&232.841&2.7&10.7235&GFK70&Sumita\\
12&-123.237&1&10.7408&E-LAF11&Hikari\\
13&24.361&7.3&10.8094&LITHO-CAF2&Schott\\
14&-17.837&1&10.9498&&\\
15&40.318&1&10.1101&N-KZFS8&Schott\\
16&11.663&8.3&9.2935&LITHO-CAF2&Schott\\
17&-18.121&0.2&9.3089&&\\
18&12.026&1.2&7.7885&S-LAH63&Ohara\\
19&8.972&6&6.9603&LITHO-CAF2&Schott\\
20&-23.203&0.9&6.3056&N-SK14&Schott\\
21&19.497&0.2&5.5824&&\\
22&6.568&5&5.0979&E-FKH1&Hikari\\
23&-33.082&2.5&3.8578&S-LAH66&Ohara\\
24&4.005&3&2.4296&&\\
25&-4.235&2&2.1726&N-LAK33B&Schott\\
26&8.775&3.3&2.668&J-SF03&Hikari\\
27&-10.282&&3.0054&&\\\hline
\end{supertabular}

\topcaption{60x, 1.4NA Objective from Olympus Patent~\cite{suzuki1996Immersion}}
\label{olympus1p4_60x}
\begin{supertabular}{ c c c c c c }
\hline
\underline{\smash{Surf       }}&\underline{\smash{r(mm)}}&\underline{\smash{d(mm)}}&\underline{\smash{h(mm)}}&\underline{\smash{Glass}}&\underline{\smash{Mfr}}\\
1&Infinity&0.17&0.221&1.521000, 56.020000&\\
2&Infinity&0.14&0.778&1.515480, 43.100000&\\
3&Infinity&0.6&0.73811&BSL7&Ohara\\
4&-1.8192&3.84&1.1653&LAH58&Ohara\\
5&-3.2177&0.1&3.2172&&\\
6&-20.4857&2.1418&4.7439&N-PSK58&Schott\\
7&-8.7588&0.3&5.1808&&\\
8&11.0685&5.3&6.3291&FPL51&Ohara\\
9&-10.4406&1&6.2509&BPM4&Ohara\\
10&18.9938&4.5&6.237&FPL53&Ohara\\
11&-17.4921&0.15&6.3246&&\\
12&25.511&1&6.0815&BPH40&Ohara\\
13&6.4981&6.5&5.5195&FPL53&Ohara\\
14&-16.9602&1&5.638&BPH50&Ohara\\
15&-37.6734&0.3&5.7893&&\\
16&8.7662&3.1&5.9715&FPL52&Ohara\\
17&145.8837&0.15&5.7691&&\\
18&7.866&5.734&5.13&PHM52&Ohara\\
19&-8.8483&1&3.5547&BPH40&Ohara\\
20&3.0648&3.2&2.2843&&\\
21&-3.4631&2.0409&2.1276&BPH50&Ohara\\
22&270.3729&6.7011&2.8398&TIH6&Ohara\\
23&-8.4836&&4.2064&&\\\hline
\end{supertabular}

\topcaption{100x, 1.65NA Objective from Olympus Patent~\cite{suzuki1997Immersion}}
\label{olympus1p65_100x}
\begin{supertabular}{ c c c c c c }
\hline
\underline{\smash{Surf       }}&\underline{\smash{r(mm)}}&\underline{\smash{d(mm)}}&\underline{\smash{h(mm)}}&\underline{\smash{Glass}}&\underline{\smash{Mfr}}\\
1&Infinity&0.17&0.11&S-YGH52&\\
2&Infinity&0.1289&0.81&1.780350, 19.0701&\\
3&Infinity&0.51&0.7276&S-YGH52&Ohara\\
4&-3.437&2.17&1.3207&LAH58&Ohara\\
5&-2.2093&0.1325&2.2008&&\\
6&-10.9949&2.45&3.668&LAH58&Ohara\\
7&-5.8271&0.1997&4.3677&&\\
8&12.675&5.32&5.3696&FPL53&Ohara\\
9&-22.9089&1.2&5.5172&LAL8&Ohara\\
10&10.1935&7.3&5.7944&FPL53&Ohara\\
11&-9.0192&0.2&6.4719&&\\
12&8.0162&4.8&6.0812&PHM52&Ohara\\
13&-20.6259&1.25&5.5464&BPH50&Ohara\\
14&5.2036&1&4.0885&&\\
15&5.6171&5.4&4.3743&FPL53&Ohara\\
16&-6.1286&1.2&3.9893&BPH35&Ohara\\
17&-28.4328&0.2334&3.8486&&\\
18&8.1214&4.9&3.5963&FPL53&Ohara\\
19&-13.9811&2.6848&2.5328&BPM4&Ohara\\
20&6.8433&1.3&1.8288&&\\
21&-2.6403&3.0112&1.7564&BPH35&Ohara\\
22&14.3617&2.54&2.6551&S-TIH6&Ohara\\
23&-7.4872&&2.9716&&\\\hline
\end{supertabular}

\topcaption{100x, 1.45NA Objective from Zeiss Patent~\cite{matthae2003High}}
\label{zeiss1p45_100x}
\begin{supertabular}{ c c c c c c }
\hline
\underline{\smash{Surf       }}&\underline{\smash{r(mm)}}&\underline{\smash{d(mm)}}&\underline{\smash{h(mm)}}&\underline{\smash{Glass}}&\underline{\smash{Mfr}}\\
1&Infinity&0.17&0.125&1.52216, 58.50&\\
2&Infinity&0.12&0.675&1.51536, 41.36&\\
3&Infinity&0.6&1&N-BK7&Schott\\
4&-1.2579&2.5&1.08&N-LASF31&Schott\\
5&-2.778&0.1&2.77&&\\
6&-6.5423&3.64&4.345&N-FK51&Schott\\
7&-4.9407&0.1&4.94&&\\
8&-44.666&3.09&7.15&N-PK51&Schott\\
9&-11.14&0.103&7.45&&\\
10&-139.25&1.11&7.92&N-KZFS4&Schott\\
11&12.23&7.76&8.53&N-PK51&Schott\\
12&-13.925&0.1&8.77&&\\
13&21.754&4.87&8.25&CAF2&Infrared\\
14&-15.961&1.16&8.035&N-KZFS4&Schott\\
15&9.8584&4.87&7.365&N-PK51&Schott\\
16&-51.958&0.205&7.315&&\\
17&6.7777&5.83&6.5&N-PK51&Schott\\
18&-93.056&1.1&5.915&N-KZFS4&Schott\\
19&3.5485&1.306&3.37&&\\
20&4.5281&4.85&3.425&N-PK51&Schott\\
21&3.759&1.907&2.255&&\\
22&-3.1612&1.92&2.235&SF2&Schott\\
23&-2.8175&0.73&2.6&N-FK51&Schott\\
24&-8.1748&&2.915&&\\\hline
\end{supertabular}



\end{document}